# A Computationally Efficient and Practically Feasible Two Microphones Blind Speech Separation Method

Chandan K A Reddy, Gautam S Bhat, Nikhil Shankar, Issa Panahi

*Abstract*— Traditionally, Blind Speech Separation techniques are computationally expensive as they update the demixing matrix at every time frame index, making them impractical to use in many Real-Time applications. In this paper, a robust data driven two-microphone sound source localization method is used as a criterion to reduce the computational complexity of the Independent Vector Analysis (IVA) Blind Speech Separation (BSS) method. IVA is used to separate convolutedly mixed speech and noise sources. The practical feasibility of the proposed method is proved by implementing it on a smartphone device to separate speech and noise in Real-World scenarios for Hearing-Aid applications. The experimental results with objective and subjective tests reveal the practical usability of the developed method in many real-world applications.

## I. INTRODUCTION

Separating sources that are convolutedly mixed is one of the most challenging problems to solve. Separating speech source from spatially mixed noise [1] is gaining lot of attention as it finds numerous applications in the widely used platforms such as hearing aids, smartphones, home assistant devices, entertainment and multimedia. Most of these devices come with at least two microphones, allowing us to use temporal, spectral and spatial properties of the signals, as opposed to only temporal and spectral characteristics available in a single microphone case. In the recent times, unsupervised (blind) algorithms are widely used, which utilize the statistical independence and higher order statistics for source separation. For convolutive mixtures, frequency-domain methods are preferred over time-domain approaches due to the computational efficiency [1, 2]. Independent Vector Analysis (IVA) is shown to be an effective Blind Speech Separation (BSS) method for convolutive mixtures as it inherently solves the permutation problem that many other frequency-domain approaches face [3-6]. IVA is an iterative approach to estimate the demixing matrix. To apply IVA for applications that demand Real-Time operations, on-line versions of IVA with faster convergence speeds are developed [7-10]. The Natural gradient based online methods were initially developed whose performance depend on the choice of step size. Auxiliary function based IVA (AuxIVA) is a BSS method that does not depend on any environmental sensitive parameters such as step size and it is shown to have faster convergence, but with a little compromise in separation performance [9-12].

Recently, smartphones are used as auxiliary computing devices to Hearing Aids (HA) [13-15]. The noisy speech is captured using the microphones on the smartphone, processed using the smartphone's processor and the output is wirelessly transmitted to the HA. Applying BSS for this kind of application will enhance the speech, thereby improving the perceptual audio quality for hearing impaired. However, the existing online IVA methods are limited to work on smartphones due to the iterative nature of computing the demixing matrix for each incoming frame, which is referred to as the algorithmic delay. Computing the demixing matrix for every frame is redundant, as the acoustic conditions do not change in short intervals in real-world conditions. For Real-Time applications, the algorithmic delay should be less than the frame length [16] for flawless operations. Although, many Real-Time online IVA methods have been proposed over the years [7, 11], the algorithmic delays of these methods are higher than the frame lengths making them unsuitable to run on the smartphones for real-time playback. In [17], a state of the art time-domain implementation of online IVA is proposed to reduce the algorithmic delay. They compute the demixing matrix for each incoming frame, which is still redundant and can be further improved using our approach.

In this contribution, we propose a criterion to update the demixing matrix, which greatly reduces the average computational complexity required to separate the sources. The demixing matrix in the frequency domain is the impulse response between the sources and the microphones in the time domain. The proposed criterion is based on tracking significant changes in the impulse response between the sources and the microphones. The demixing matrix is updated only at the frame (time index) where the proposed criterion is satisfied, rather than updating at every frame. This approach exponentially reduces the average number of computations required to obtain the demixing matrix, making the online-based IVA methods suitable for Real-Time applications. The proposed criterion depends on a computationally fast two-microphone speech source Direction of Arrival (DOA) estimation. Over the years, several DOA estimators are developed to work in noisy and reverberant conditions [18, 19]. Among the broad categories of DOA estimators, algorithms based on Time Delay of Arrival (TDOA) using the Generalized Cross Correlation (GCC) [20] are known to be computationally efficient. However, these methods are more susceptible to different kinds of

This work was supported by the National Institute of the Deafness and Other Communication Disorders (NIDCD) of the National Institutes of Health (NIH) under Award 5R01DC015430-03. The content is solely the responsibility of the authors and does not necessarily represent the official views of the NIH.

environmental noises and yields erroneous estimate of DOA, which makes them unsuitable to track source location changes to reduce the computations in IVA. We propose a Feed Forward Neural Network (FNN) based DOA estimation (FNNDOA) of speech source in noisy and reverberant environment. The proposed method is computationally efficient and outperforms traditional techniques in terms of estimation accuracy. The integrated setup of FNNDOA and online IVA is implemented on a smartphone that seamlessly works as an assistive device to the Hearing Aids to prove the concept. The algorithmic delay of the entire pipeline is less than the frame length, enabling it to run flawlessly. The integrated setup is evaluated objectively using BSS and speech quality metrics. We also performed subjective tests and received positive feedback.

## II. TDOA BASED DOA ESTIMATION

In the literature, the DOA estimation algorithms are broadly categorized based on their signal processing procedure to obtain the estimate of DOA. Some of the categories are Subspace based approaches [21-23], TDOA based techniques using GCC and Least Squares (LS) [24, 25], model based approaches such as the Maximum Likelihood [26], and methods based on blind identification of the impulse response between the source and the microphones [27, 28]. Among these methods, there is always a tradeoff between the computational cost and the accuracy of the algorithm in estimating the DOA. In this work, the intention of using DOA is to reduce the computational complexity of the online IVA to make it feasible to run on the smartphone. TDOA based DOA estimators using GCC are known to be computationally very fast. However, the TDOA methods are not robust in the presence of noise, due to the impractical assumptions on the signal model. The original GCC framework [24] is based on the single path plane wave propagation of sound waves from a single sound source that is received at the two microphones $x_1(n)$ and $x_2(n)$ that are sufficiently separated. The GCC function to calculate the delay is given by,

$$\hat{\gamma}_{x_1 x_2}(m) = E[x_1(n) x_2(n-m)] \quad (1)$$

where $E$ denotes expectation, $\widehat{(.)}$ denotes the estimated value and $m$ is a dummy variable. The argument $\eta$ that maximizes (1) is the estimated TDOA given by,

$$\hat{\eta} = \arg\max \hat{\gamma}_{x_1 x_2}(\eta) \quad (2)$$

The estimated DOA angle $\hat{\theta}$ is given by,

$$\hat{\theta} = \cos^{-1} \frac{\hat{\eta} c}{d} \quad (3)$$

where $c = 343$ m/s is the speech of sound, $d$ (in meters) is the distance between the microphones and $\hat{\eta}$ is in seconds.

The DOA estimate using (3) will be accurate only in the noise free conditions or when the noise at the 2 mics are uncorrelated. In reality, these conditions do not hold true due to the presence of various kinds of noises and reverberation resulting in erroneous $\hat{\theta}$. Hence, the peak detection using the GCC function in (1) will erroneously point to noise or to the directions of reflections in the presence of reverberation. This makes it impractical for IVA application.

## III. PROPOSED FNN BASED DOA ESTIMATION

Let $\boldsymbol{r}(n) = [|r_{x_1 x_2}(-m)|, \ldots, |r_{x_1 x_2}(0)|, \ldots, |r_{x_1 x_2}(m)|]$ be a vector of absolute values of the cross-correlation between the input frames from the two microphones $x_1$ and $x_2$ at time index $n$. We drop $n$ for brevity. The input feature vector is composed of the normalized cross-correlation coefficients of $\boldsymbol{r}$ at the valid lags $-m$ to $m$ given by,

$$\boldsymbol{U} = \frac{\boldsymbol{r}(n)}{\max(\boldsymbol{r})} \quad (4)$$

In the case of 16 kHz sampling rate and 13 cm separation between the microphones (smartphone scenario), the value of $m$ is 6, which is the maximum delay in samples between the microphones. Any value above $m = 6$ will be invalid to estimate the DOA. A FNN with one hidden layer consisting of 8 nodes is used to capture the non-linear relationship between $\boldsymbol{U}$ and the DOA. A "Rectified Linear Unit (ReLu)" is chosen as the activation function at the nodes of the hidden layer. Let $\boldsymbol{Z}_1$ denote the output of the hidden layer given by,

$$\boldsymbol{Z}_1 = \max(0, \boldsymbol{W}_1 \boldsymbol{U}) \quad (5)$$

$\boldsymbol{W}_1$ is the linear transformation weights from the input layer to the hidden layer. Max function is used to introduce non-linearity in the hidden layer, which is also the ReLu activation function. This helps in learning non-linear relationship between the inputs and the outputs. Let $\boldsymbol{Z}_2 = \boldsymbol{W}_2 \boldsymbol{Z}_1$ be the linear transformation of $\boldsymbol{Z}_1$ to the output. $\boldsymbol{W}_2$ is the weights of the connections from the hidden layer to the output nodes. The weight vectors $\boldsymbol{W}_1$ and $\boldsymbol{W}_2$ is obtained using the first-order gradient-based optimization of stochastic objective function, which is called as 'Adam' [29]. The output layer of the Neural Network consists of the output classes, which are the angles of the DOA. In this work, 7 different angles between $0^0$ and $180^0$ are considered with a separation of $30^0$. The explanation for choosing only 7 of these angles will be given in the next section. Softmax function is used at the output nodes to give the probabilities of each class given by,

$$p(\theta_n = c | \boldsymbol{U}(n)) = \frac{\exp(\boldsymbol{Z}_2(n))}{\sum_{k=1}^{C} \exp(\boldsymbol{Z}_2(n))}, c \in (0, C-1) \quad (6)$$

Each of the output class $c$ will have a probability associated to it and the one with the highest probability will be the most probable class. $C$ is the number of output classes.

## IV. FNNDOA USED AS AN APPLICATION TO IVA

### A. Frequency Domain BSS model

The mixing process in the real-world acoustic environment includes delays, attenuations and reverberation, which can be well represented by the convolution mixing process. If there are $P$ sources and $Q$ sensors, the signal captured by sensor $q$ is given by,

$$x_q(n) = \sum_{p=1}^{P} a_{qp}(n) * s_p(n) \quad (7)$$

where $(Q \geq P)$, $(*)$ is the convolution. $a_{qp}(n)$ is the finite duration impulse response mixing filter from source $p$ to sensor $q$ is given by,

$$x_q^{[f]}(m) = \sum_{p=1}^{P} a_{qp}^{[f]} * s_p^{[f]}(m) \quad (8)$$

$$\boldsymbol{x}^{[f]}(m) = \boldsymbol{A}^{[f]} \boldsymbol{s}^{[f]}(m) \quad (9)$$

where $s_p^{[f]}(m)$, $x_q^{[f]}(m)$ and $a_{qp}^{[f]}$ are frequency domain signals of $s_p(n)$, $x_q(n)$ and $a_{qp}(n)$ respectively at frame index $m$.

$x^{[f]}(m) = \left[ x_1^{[f]}(m), x_2^{[f]}(m), \ldots \ldots, x_Q^{[f]}(m) \right]$, $s^f(m) = [s_1^{[f]}(m), s_2^{[f]}(m), \ldots \ldots, s_P^{[f]}(m)]$ and $A^{[f]}$ is the mixing matrix for frequency bin $f$, with $a_{qp}^{[f]}$ as its entries for each frame $m$. The goal of IVA is to find a demixing matrix $W^{[f]}$ at each frequency bin $f$ such that,

$$y^{[f]}(m) = W^{[f]} x^{[f]}(m) \quad (10)$$

where $y^{[f]}(m)$ is the estimate of $s^{[f]}(m)$.

In (10), the problem can be viewed as estimating the demixing matrix using Independent Component Analysis (ICA) at each frequency bin. When using algorithms like ICA, permutation problem should be carefully addressed, otherwise, the separation of the sources fails. IVA on the other hand makes use of inter-frequency bin information to solve the permutation problem. The only difference between ICA and IVA is that, the signals are considered as vectors instead of scalars, and they will be optimized as multivariate variables instead of univariate variables. Using Kullback-Leiblar divergence as the objective function, the update equation for the demixing matrix is optimized using gradient descent which is given by,

$$W^{[f]} = W^{[f]} + \eta \ \{I - \mathrm{E} \ [\Phi^{[f]}(y^{[f]})(y^{[f]})^H]\} \ W^{[f]} \quad (11)$$

Where $\Phi^{[f]}(y_p) = \dfrac{y_p^{[f]}}{\sqrt{\Sigma_{f=1}^{F} |y_p^{[f]}|^2}}$ is a non-linear function that is typically shown to give better separation [17]. In (11), $y^{[f]}$ is calculated from the previous frame to obtain the demixing matrix for the current frame.

*B. Proposed integration of FNNDOA and IVA to achieve computational efficiency*

The proposed method is shown in Figure 2 (a). The DOA is estimated using FNNDOA method on the voice only frames which are detected using a voice activity detector (VAD) [14]. Then it checks for the condition that is described in Figure 2 (b). The demixing matrix is only updated if the position of the speech source changes by at least $30^0$. The noise is assumed to be diffused. Otherwise, it uses the old demixing matrix for the current frame. The proposed condition is used to decrease the number of false positive detection of change in DOA of $30^0$. If the frames between index $(n-5)$ to $(n-1)$ have a DOA estimate of $\theta_2$, and the frames with indices $(n-10)$ to $(n-6)$ have a different DOA estimate say $\theta_1$, then we say that the condition is satisfied. We relax the condition by enforcing the requirement as 4 out of 5 frames to have same DOA instead of 5 out of 5. This is to avoid the percentage of true negatives as the DOA estimate will not be 100% accurate. If we go anything lower than 4, the accuracy of correctly detecting the change in DOA will decrease. The resulting demixing matrix is then used to separate the sources using (10). The time domain signal of the separated sources is obtained using Inverse Fast Fourier Transform (IFFT).

*C. Smartphone implementation of the method*

In this work, Google Pixel running Android 7.1 Nougat operating system is used as an assistive device to Hearing Aids (HA). We use two microphones on Google Pixel, separated by 13 cm to capture the audio data. The noisy signal captured is processed and wirelessly transmitted to the HA. The developed code can also run perfectly in other Android smartphones. The smartphone considered has an M4/T4 HA Compatibility rating and meets the requirements set by Federal Communications Commission (FCC). The proposed BSS setup that depends on FNNDOA in implemented to work in real-time on the smartphone using Android Studio [30]. Figure 3 shows the snapshot of the app that has various controls to turn on and off the BSS. We can also choose the desired source for playback through the HA. An inbuilt android audio framework was used to carry out dual microphone input/output handling. The input data acquired at 48Khz sampling frequency is down sampled to 16 KHz. A 20 ms frame was considered with 50% overlap. Since we operate in frequency domain, the FFT size was set to be 512 for the input buffer. For real-time audio processing, lowest hardware permitted audio latency is needed to avoid any skipping of input frames. Thus, the separation should take place within the time of audio input-output frame, i.e. within 320 samples or 20ms for Android-based smartphones. The traditional IVA method that updates demixing matrix at every frame cannot be computed within 20 ms as it requires many iterations to converge. The proposed method is able to run without skipping frames and separate the sources. The experimental results are discussed in the next section.

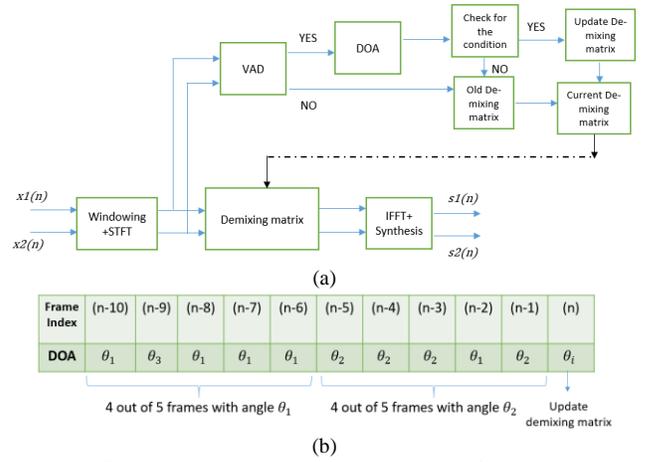

Fig. 2. (a) Block diagram of the proposed method, (b) Proposed condition to update demixing matrix

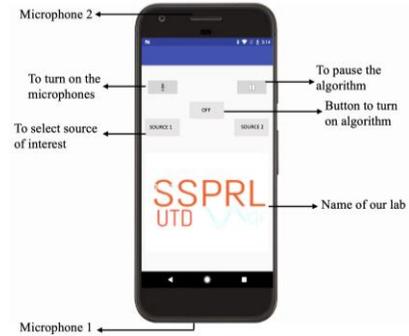

Fig 3 Snapshot of the developed smartphone application

V. EXPERIMENTAL RESULTS

*A. Dataset to evaluate the proposed method*

The data for assessing the performance of the proposed method using objective measures was generated using the Image Source Method (ISM) toolbox [31]. The clean speech was used from the TIMIT and HINT database. The clean speech source was placed at different angles to the microphone pair. The angles considered for the speech source direction were [$0^0, 45^0, 90^0, 135^0$ and $179^0$]. The noise is assumed to be diffused. The speech and the noise files were sampled at 16 kHz and the noisy speech files were generated using ISM toolbox. The distance between the two microphones is 13 cm. The distance between the speech source and the microphones is 2.5 m. The noisy speech files with a frame size of 20 ms is used for processing the IVA with 50% overlap.

### B. Objective and Subjective Evaluation:

The proposed method is evaluated using a performance for source separation, Signal to Distortion Ratio (SDR) [32] and PESQ [33] for speech quality. Figures 4 and 5 shows the performance evaluation plots using the above-mentioned objective measures for speech mixed in Babble and Machinery noise types at SNR levels of -5 dB, 0 dB and 5 dB. The proposed method is compared with noisy speech, dual microphone spectral-coherence based SE method [34], and BSS using the traditional batch-wise IVA which is in non-real time. The BSS using the proposed setup outperforms noisy and spectral-coherence method in terms of all measures. The performance of the proposed method is on par with the traditional IVA. However, there is a tradeoff in computational time and the accuracy of the proposed method in comparison to the traditional IVA approach.

Since the proof is in the pudding, we conducted subjective tests using Mean Opinion Scores (MOS) [35]. We performed MOS tests on 10 expert normal hearing subjects who were presented with noisy speech, enhanced speech using the spectral-coherence, proposed and batch wise IVA methods at SNR levels of -5 dB, 0 dB and 5 dB. For each audio file the subjects were instructed to score in the range of 1 to 5 with 5 being excellent speech quality and 1 being bad speech quality. They were given the flexibility to go back and change the score as well. The detailed description of scoring procedure is in [35]. Subjective test results in Figure 6 illustrates the effectiveness of the proposed method in separating the two sources.

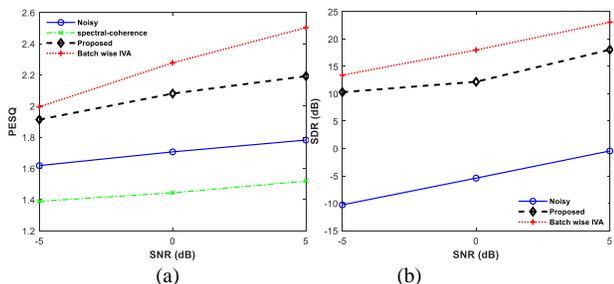

Fig. 4 Performance evaluation of speech mixed with Babble Noise using (a) PESQ, (b) SDR

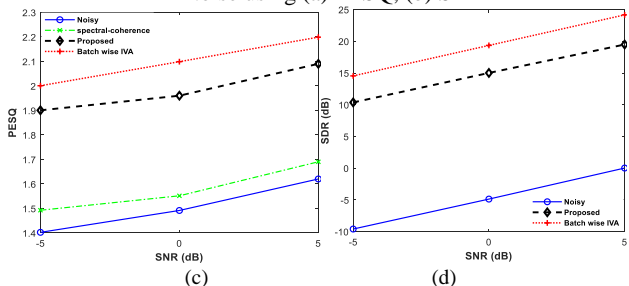

Fig. 5 Performance evaluation of speech mixed with Machinery Noise using (a) PESQ, (b) SDR

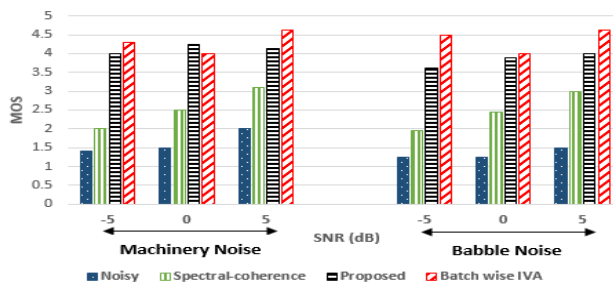

Fig. 6 Subjective test results

It is worth noting from the results that the proposed method cannot outperform the traditional batch wise IVA, except for one condition in Machinery SNR 0 dB case in subjective results. This might probably be due to a bias in the perceptual preference of few subjects. Overall, there is a tradeoff between computational efficiency and performance. But the difference in the performance of the proposed and traditional approach is not significant in comparison to the improvement over noisy. The sample audio files can be found in [36]. The proposed setup can also be expanded to other variations of IVA [5-12].

### C. Computational Complexity of the Proposed method:

The computational complexity of the proposed method can be analyzed by calculating the number of times the demixing matrix is updated for a fixed length of noisy speech. For instance, let us consider a noisy speech file of 15 secs with a sampling rate of 16 kHz in which the source direction changes every 3 secs. Therefore, according to the proposed criterion in Figure 2 (b), the changes in the source direction will be detected 4 times. Even if a larger frame of length (say 100 ms) is processed without any overlap, a total of 150 frames should be processed. The traditional online-based IVA computes demixing matrix 150 times. On the other hand, the proposed method updates the demixing matrix only 4 times in the entire 15 secs of the data. Hence, for this scenario, the proposed method is 37 times computationally efficient than the traditional approach. The computational time of the proposed method depends on the number of times the change in the source location is detected. In realistic scenarios, the speaker does not change the position often while talking. But there are instances where speaker do change position while talking such as a person presenting on a podium. In such cases, the computational complexity of the proposed method will converge to that of the traditional IVA. Hence, the computational complexity of the proposed method will be less than or equal to the traditional IVA method.

## VI. CONCLUSION

A computationally efficient BSS technique based on IVA was proposed. A FNN based DOA estimate is used to reduce the average number of computations required to separate the speech source from noise. Overall, the proposed method gives impressive results and works seamlessly on a smartphone thereby proving the concept.